\documentclass{elsart}
\usepackage{graphicx}
\usepackage{natbib}

\begin{document}
\def\aj{AJ }%
\def\araa{ARA\&A }%
\def\apj{ApJ }%
\def\apjl{ApJ }%
\def\apjs{ApJS }%
\def\ao{Appl.~Opt. }%
\def\apss{Ap\&SS }%
\def\aap{A\&A }%
\def\aapr{A\&A~Rev. }%
\def\aaps{A\&AS }%
\def\azh{AZh }%
\def\baas{BAAS }%
\def\jrasc{JRASC }%
\def\memras{MmRAS }%
\def\mnras{MNRAS }%
\def\pra{Phys.~Rev.~A }%
\def\prb{Phys.~Rev.~B }%
\def\prc{Phys.~Rev.~C }%
\def\prd{Phys.~Rev.~D }%
\def\pre{Phys.~Rev.~E }%
\def\prl{Phys.~Rev.~Lett. }%
\def\pasp{PASP }%
\def\pasj{PASJ }%
\def\qjras{QJRAS }%
\def\skytel{S\&T }%
\def\solphys{Sol.~Phys. }%
\def\sovast{Soviet~Ast. }%
\def\ssr{Space~Sci.~Rev. }%
\def\zap{ZAp }%
\def\nat{Nature }%
\def\iaucirc{IAU~Circ. }%
\def\aplett{Astrophys.~Lett. }%
\def\apspr{Astrophys.~Space~Phys.~Res. }%
\def\bain{Bull.~Astron.~Inst.~Netherlands }%
\def\fcp{Fund.~Cosmic~Phys. }%
\def\gca{Geochim.~Cosmochim.~Acta }%
\def\grl{Geophys.~Res.~Lett. }%
\def\jcp{J.~Chem.~Phys. }%
\def\jgr{J.~Geophys.~Res. }%
\def\jqsrt{J.~Quant.~Spec.~Radiat.~Transf. }%
\def\memsai{Mem.~Soc.~Astron.~Italiana }%
\def\nphysa{Nucl.~Phys.~A }%
\def\physrep{Phys.~Rep. }%
\def\physscr{Phys.~Scr }%
\def\planss{Planet.~Space~Sci. }%
\def\procspie{Proc.~SPIE }%
\let\astap=\aap
\let\apjlett=\apjl
\let\apjsupp=\apjs
\let\applopt=\ao

\begin{frontmatter}
\title{The influence of grain rotation on the structure of dust
  aggregates}

\author[Ams]{Paszun D},
\author[Ams]{Dominik C.}

\address[Ams]{Astronomical Institute "Anton Pannekoek", 
University  of Amsterdam, Kruislaan 403 1098 SJ Amsterdam, 
The Netherlands }

\begin{abstract} 
We study the effect of rotation during the collision between dust
aggregates, in order to address a mismatch between previous model
calculations of Brownian motion driven aggregation and experiments.
We show that rotation during the collision does influence the shape 
and internal structure of the aggregates formed.  The effect is 
limited in the ballistic regime when aggregates can be considered 
to move on straight lines during a collision.  However, if the 
stopping length of an aggregate becomes smaller than its physical 
size, extremely elongated aggregates can be produced. We show that 
this effect may have played a role in the inner regions of the solar 
nebula where densities were high.
\end{abstract}
\end{frontmatter}

\section{Introduction}
Dust aggregation plays a central role in most theories of planet
formation. While for the formation of giant planets, disk 
instability scenarios continue to be discussed 
\citep{1997Sci...276.1836B}, dust aggregation certainly stands at 
the beginning of the formation of planetesimals and therefore of 
the terrestrial planets.  Independent of the detailed process that 
finally forms planetesimals \citep[e.g.][]{1980Icar...44..172W,
2002ApJ...580..494Y,2001ApJ...546..496C}, dust grains first need to 
grow and settle towards the midplane of the disk.  Planet formation 
therefore starts with the first steps of dust aggregation, when 
dust grains inherited by the disk from the interstellar medium 
start to collide and grow.  This first step of growth is governed 
by Brownian motion aggregation.  In the high density regions of the 
disk, dust/gas coupling is so tight that the main source of relative 
velocities between grains are the random motions produced by 
individual collisions between gas atoms and the grains. Understanding 
the Brownian motion phase of dust aggregation therefore means 
understanding the first step towards planets in disks.

At the very low collision velocities produced by Brownian motion
($\sim$cm/s), grains always stick and no restructuring is occurring 
in the collision \citep{1997ApJ...480..647D,2000Icar..143..138B}. The
structure of aggregates formed in this regime is therefore indeed a
pure probe of the physical processes driving growth with Brownian
motion of the particles.  Theoretically, growth by Brownian motion 
was studied for example by \citet{1999Icar..141..388K}. They 
calculated the motion of micron sized dust grains enclosed in a box 
of approximately constant dust number density.  Diffusion, caused by 
the presence of a gaseous medium, produces relative motion of grains 
and leads to the growth of dust.  The orientation of the (spherically 
not symmetric) aggregates was not followed during the computations.  
Instead, in order to randomize the relative orientation during 
collisions, the orientations of collision partners just before a 
collision was selected randomly. With these initial conditions, the 
aggregates were left to collide, without considering rotation 
\emph{during} the collision.  The calculations show a slow growth of 
the dust aggregates with time.  At any time, the box contains a 
distribution of aggregate shapes which is characterized by a 
distribution of fractal dimensions.  The mean fractal dimension
found in the numerical study is around $D_f = 1.8$.

On the experimental side, a series of low-gravity experiments has
been conducted \citep{2000PhRvL..85.2426B,2004PhRvL..93b1103K} in
order to study the growth of fractal aggregates under Brownian 
motion conditions.  The results confirm many expected aspects of 
this growth regime, but also showed unexpectedly elongated 
aggregates, in contrast with the predictions by 
\citet{1999Icar..141..388K}.  

The authors qualitatively argue that rotation of the aggregates 
during the collision might
lead to more elongated aggregates, because the probability of
achieving contact between two aggregates at larger separations
increases if the aggregates rotate.

To compare experimental and theoretical results, it is necessary to
quantify the visual impression of \emph{elongation}.  In this study 
we will be using two different ways to do so:

\begin{enumerate}
\item\label{item:1} One can define an \textbf{elongation factor} as
  the ratio of maximum to minimum diameter of aggregate
\begin{equation}
\label{eq:1}
f_\mathrm{el}=\frac{d_\mathrm{max}}{d_\mathrm{min}}
\end{equation}
where $d_\mathrm{min}$ is measured in perpendicular direction to
$d_\mathrm{max}$.  Experimentally, this quantity has to be derived
from a few, or even a single picture.  When measuring this value 
for model aggregates, we therefore choose three different 
projections of the aggregate and determine $d_\mathrm{max}$ and 
$d_\mathrm{min}$ from these images.

\item We can also derive the \textbf{fractal dimension} of the
  aggregates.  This quantity can be computed by measuring the
  mass--radius relation of aggregates of different mass $m$ and 
  size $R$. If that relation follows a power law
\begin{equation}
m \propto R^{D_\mathrm{f}}
\label{D_f}
\end{equation}
then $D_\mathrm{f}$ is called the fractal dimension of the
aggregate.  The fractal dimension measured in the experiments 
is 1.4, and typical values predicted by the numerical calculations 
are around 1.8, indicating a mismatch.
\end{enumerate}

While the idea of particle rotation is appealing as a mechanism to
increase the elongation of dust aggregates, this has never been 
shown quantitatively.  In this paper we investigate the influence 
of rotation on the structure of aggregates formed under Brownian 
motion conditions.

\section{The model}
In order to test the influence of rotation on the growth of 
aggregates we developed a model for collisions of rotating 
aggregates. Since we are interested in the low velocity regime, 
collisions are not energetic enough to cause restructuring.  These 
processes are therefore neglected, and aggregates are treated as 
rigid entities.  Our code calculates the equation of motion of two 
aggregates set on collision course with each other.  When both 
aggregates get into the first contact between any two constituent 
grains, a new aggregate is formed.  The structure of this aggregate 
is determined by the position and orientation of the two aggregates 
at the moment of contact -- no further changes occur after this 
moment.  

In all our calculations we used the same monomers sizes and masses 
as in the CODAG experiment~\citep{2004PhRvL..93b1103K}. Each monomer 
was $1 \mu$m in diameter and its mass was $1.0\times10^{-15}$ kg. 
The temperature assumed in our calculations was $300$K , again like 
in the experiment.

In a fully general calculation, one would have to consider an ensemble
of particles and follow the growth of this ensemble.  However, it has
been shown experimentally that the mass distribution function during
Brownian motion growth is very narrow, so that a mono-disperse
approximation works very well \citep{2004PhRvL..93b1103K}.  We
therefore proceeded in the following way: Starting with dimers, we
first compute a large number of collisions between dimers, with random
initial position, rotational orientation, and impact parameter.  The
translational and rotational velocities are taken from a Maxwellian
distribution for a given temperature $T$.  The resulting quadrumer
structures are stored in a database.  In the next step, we collide two
quadrumers selected randomly from the database.  A database of
quadrumer structures is built in this way.  In a similar way, we
produce aggregates with larger sizes, always containing $2^n$ grains.

\begin{figure}[!h]
\centering
\includegraphics[width=1.0\textwidth]{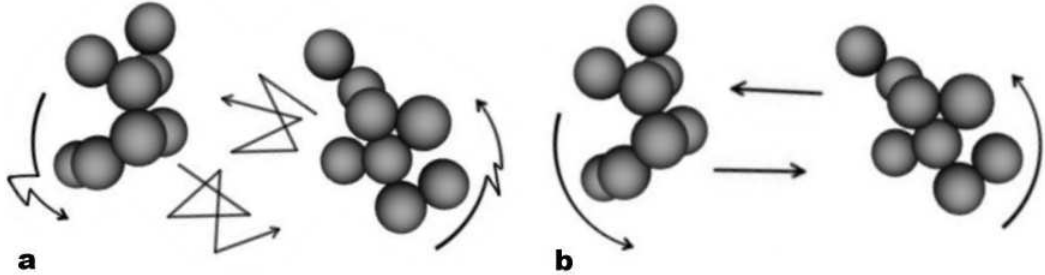}
\caption{ 
Schematic sketch of collisions between two aggregates in two
  cases. a - non--ballistic collision with mean free path of an
  aggregate shorter than the aggregate size. Velocities change every
  $\tau_\mathrm{f}$. b - ballistic collisions with mean free path
  longer than size of an aggregate. }
\label{exmpl}
\end{figure}

In order to solve the problem we have studied the influence of
rotation in two separate cases. The first one assumes \emph{ballistic
  collisions} where the mean free path of an aggregate is longer than
its size.  The second assumes aggregation in a \emph{non--ballistic
  regime}.  The stopping length in this case is shorter than the size
of the aggregate.

\subsection{Ballistic collisions}

In the ballistic regime, the mean free path of all involved particles
is larger than the size of the largest particle.  In this case, the
center of mass of each particle moves on a straight line during the
collision.  In physical terms, the ballistic limit is reached in the
limit of low (but non--zero, because the gas still needs to cause
Brownian motion of the grains) gas densities.  This is generally
assumed to be the case both in protoplanetary disks, and in
microgravity experiments described above.

When aggregates are colliding without rotation and on linear
trajectories, elongated aggregates are formed when the first contact
is made between grains on the outside of the aggregate, while the two
aggregates are more or less aligned.  Depending on the initial
orientation, this situation is realized for different impact
parameters: If the aggregates are aligned along the direction of
relative motion between the colliding aggregates, a small impact
parameter is needed.  If the alignment is perpendicular to the
collision direction, a large impact parameter produces the most
elongated result, while a small impact parameter would lead to a more
compact structure.
If the aggregates are rotating during approach, the chances that the
contact will be made early with an elongated geometry increase.  For
this process to be efficient, the linear speed of grains far away from
the center of mass should be comparable or larger than the
translational motion.  Therefore, the larger the ratio of rotational 
and translational speeds are, the greater the average
elongation of the aggregate becomes.

This can be achieved in two ways. Superthermal grain rotation does
occur in interstellar space, where it is responsible for the alignment 
of non--spherical grains with the galactic magnetic field and in this 
way produces the interstellar polarization \citep{1979ApJ...231..404P}.
With superthermal rotation, a fast circular motion is induced while
the linear velocity remains thermal.  However, superthermal rotation
relies on small forces like non--isotropic absorption and scattering
of light \citep{1996ApJ...470..551D}, or H$_{2}$ formation on specific
sites \citep{1979ApJ...231..404P}, to accelerate the grains.  The
non--thermal rotation speeds can only be sustained at extremely low
gas densities.  In fact, the densities prevailing in protoplanetary
disks are prohibitive, and superthermal rotation can be ruled out
there \citep{1993A&A...280..617O}.

On the other hand, the effective linear translational speed can be 
slowed down by embedding the grains into gas of high density, leading 
into the regime of non--ballistic collisions.

\subsection{Non--ballistic collisions}

The mean free path $l_\mathrm{b}=v_\mathrm{th}\tau_\mathrm{f}$ of a 
dust grain is the distance an aggregate with mean thermal velocity 
$v_\mathrm{th}$ can move during one friction time. The friction time
of a dust grain is
\begin{equation}
\tau_\mathrm{f} = \varepsilon \frac{m}{\sigma_\mathrm{a}}
\frac{1}{\rho_\mathrm{g}v_\mathrm{m}}
\label{tau_f}
\end{equation} 
where $m$, $\sigma_\mathrm{a}$, $\rho_\mathrm{g}$ and $v_\mathrm{m}$
are mass, aerodynamical cross section, gas density and mean thermal
velocity of the gas molecule respectively and $\varepsilon$ is a 
proportionality coefficient which is around $0.6$ 
\citep{1996Icar..124..441B}.

If $\rho_{\mathrm{g}}$ increases, the mean free path decreases
accordingly.  When the mean free path becomes similar to the largest
aggregate involved in the collision, the assumption of a ballistic
collision is no longer valid.  Instead, the particle execute a random
walk both in linear motion and in rotation.  On average, the
aggregates will spent a longer time at large distances before moving
closer together.  This effect increases the chance of creating a
contact already at large distances, i.e. a strongly elongated aggregate.
Fig.~\ref{exmpl}a presents the schematic picture of a non--ballistic 
collision.

\section{Results}
In order to investigate the influence of rotation on the production of 
elongated aggregates, we calculated growth of rotating aggregates
in the range from ballistic to highly non---ballistic collisions.

\subsection{Ballistic collisions with and without thermal rotation}

\begin{figure}[!h]
\centering
   \includegraphics[width=1.0\textwidth]{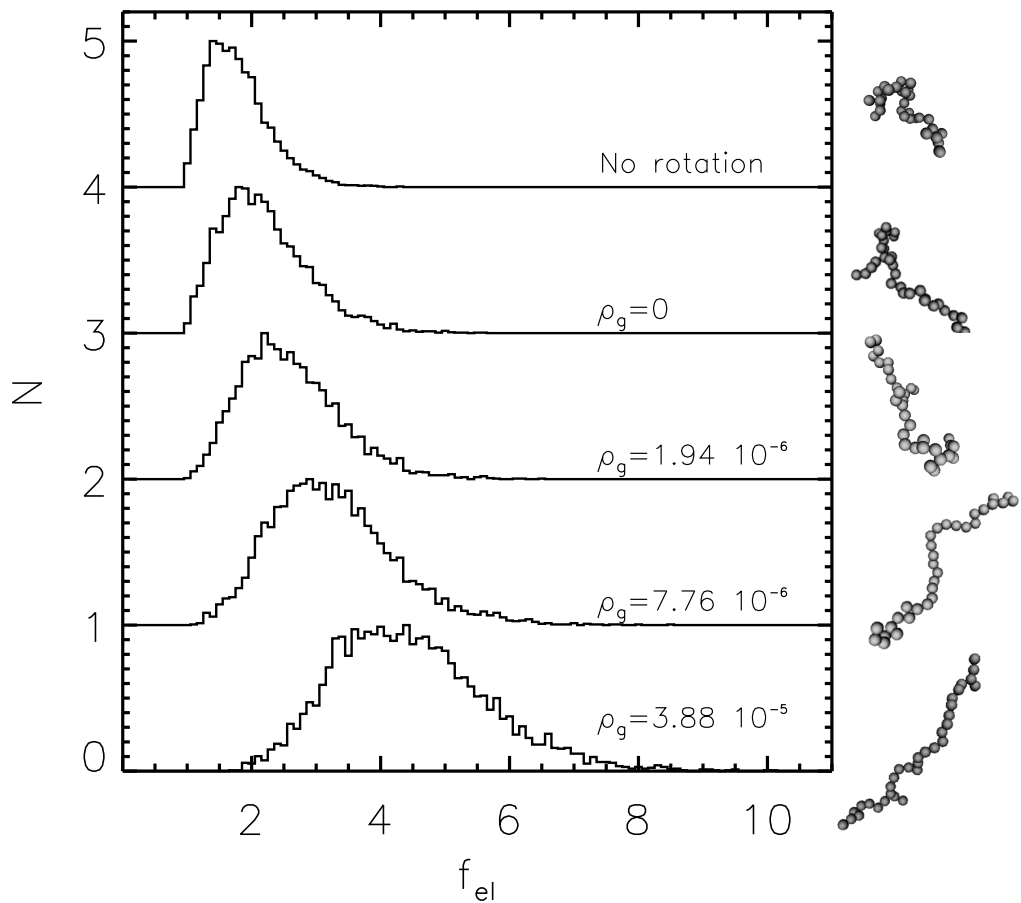}
   \caption{ 
Distribution of elongation factors, with peak value
     normalized to unity.  From top to bottom: non--rotation
     case, ballistic thermal rotation, thermal rotation for different gas
     densities $\rho_\mathrm{g}$ in $\mathrm{g\,cm}^{-3}$.  The 
     distributions are shifted vertically for better visibility.}
\label{dists}
\end{figure}

Distributions of $f_\mathrm{el}$ for ballistic collisions are plotted
in fig.~\ref{dists}.  In order to be able to compare with the results
of \citet{1999Icar..141..388K}, we also computed a case with rotation
turned of.  The two top diagrams show the non--rotating case and
thermal rotation case in the ballistic regime.  For the non--rotating
case, a very narrow distribution results.  This distribution peaks at
a value $f_{\mathrm{el}}\approx 1.6$.  For thermal rotation, the peak
value shifts to $f_{\mathrm{el}}\approx 2$ and the distribution
becomes wider.  For both cases, a very small fraction of particles
with elongation factors 4 and larger are produced, due to fortuitous
initial conditions in the computation.  Fig.~\ref{nr-rho_2x} shows the
mean elongation factors as a function of aggregates size. Each
point is plotted together with its $\pm 1 \sigma$ errorbars.
The no rotation case clearly has the lowest elongations and, as in 
the thermal rotation case, the line is almost flat for bigger sizes, 
meaning the growth affects the elongation very weakly. The elongation 
factors for rotating aggregates are shifted upwards from values of about 
$f_\mathrm{el}=1.8$ to about $f_\mathrm{el}=2.3$. The effect of rotation 
is also clearly seen in the fractal dimensions (fig.~\ref{df-ro}) which 
reach level of $D_\mathrm{f} \approx 1.7$ for non--rotating collisions. 
Thermal rotation causes the fractal dimension to drop to a value of 1.46.

\subsection{Rotating aggregates outside of the ballistic limit}

\begin{figure}[!h]
\includegraphics[width=1.0\textwidth]{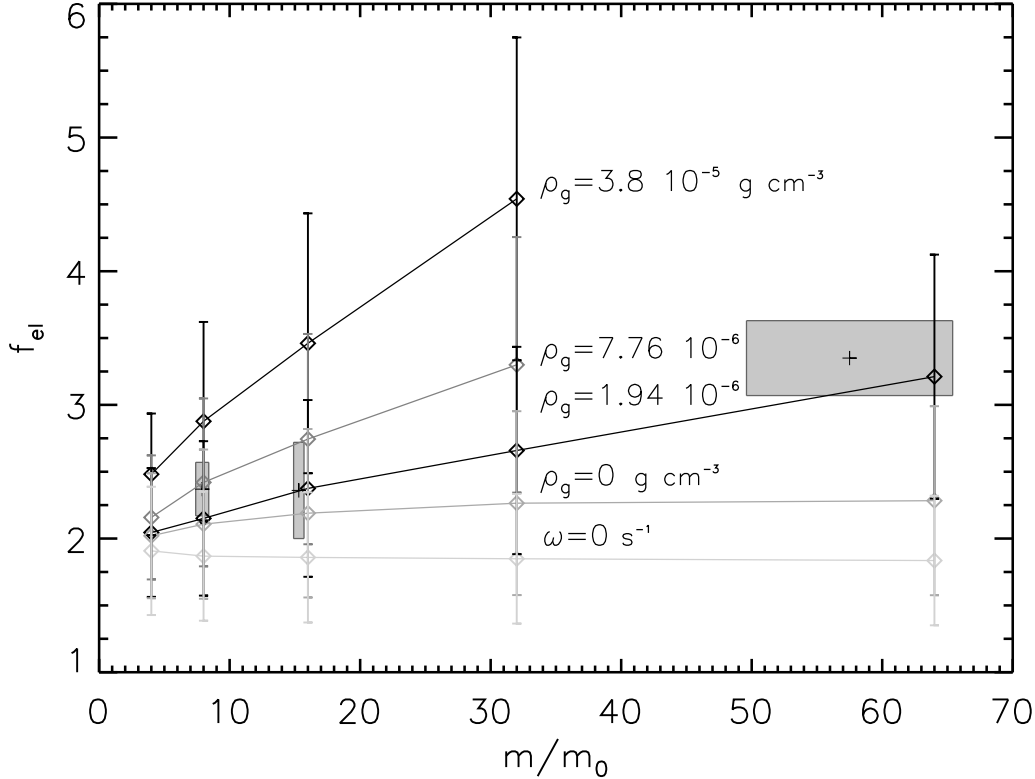}
\caption{ 
Mean elongation factor for different gas
densities. $\rho_\mathrm{g}=0$ indicates the limiting case for low
densities (mean free path much larger than aggregate sizes).
The bottom line shows the elongation factor for non--rotation case.
Three boxes are measured elongation factors of
agglomerates formed in the CODAG experiment with $\pm 1\sigma$ 
errorbars~(Blum \& Krause personal communication). }
\label{nr-rho_2x}
\end{figure}

Further calculations were done for higher gas densities.  In this
case, the friction length $l_\mathrm{b}$ becomes comparable to or
shorter than the size of the colliding aggregates.  While the thermal
velocity decreases with increasing size (mass) of an aggregate,
$\tau_\mathrm{f} \approx \mathrm{const}$ for small, non--compact 
aggregates. Consequently, the mean free path
$l_\mathrm{b}=\tau_\mathrm{f}v_\mathrm{th}$ decreases with increasing
particle size.  We performed calculations for several different gas
densities.  By tuning the gas density, we can study the transition
from the ballistic to the non--ballistic regime.  Fig.\ref{dists} shows
the resulting distribution of elongation factors for different gas
densities. Fig.~\ref{nr-rho_2x} presents the relation between the mean
elongation factor and the aggregate size for different gas densities.

The mean elongation factor $f_\mathrm{el}$ depends strongly on the
gas density $\rho_\mathrm{g}$ in the transition regime.  Increasing
the gas density from $10^{-6}$ to a few times $10^{-5}$g\,cm$^{-3}$
strongly broadens the elongation factor distribution and shifts the
peak value to $f_\mathrm{el}\approx 4$.  Extreme values of up to
$f_\mathrm{el}=10$ are reached in the tail of the distribution
function.  The elongation factor also depends on size of aggregates:
Larger aggregates reach the largest
elongations.  This is the result of the decrease of the mean free path
with increasing aggregate size. The elongation - size relation becomes
steeper for higher gas densities.
The mean elongation factor in our calculations follows a power law
dependence on the gas density.  The power index is size dependent.
Aggregates consisting of 32 monomers follow a power law with index of
0.18, while the index for those made of 16 monomers is 0.13.  This
means that the latter aggregates can reach a mean elongation
factor value about 3.4 for gas density $\rho_\mathrm{g}=3.88\times
10^{-5} \; \mathrm{g}\, \mathrm{cm}^{-3}$. For this gas density and
aggregates bigger than 32 monomers the mean elongation factor
reaches values above 4.5.

Fig.~\ref{nr-rho_2x} presents also results of the CODAG
experiment~(Blum \& Krause, personal communication).  Three points
together with their $\pm 1 \sigma$ errorbars show the ratio of maximum
to minimum diameter of aggregates formed during the experiment.  An
excellent fit with our calculations seems to be
$\rho_\mathrm{g}=1.94\times10^{-6}\; \mathrm{g}\, \mathrm{cm}^{-3}$ or
slightly higher.  The density used in the experiment was
$\rho_\mathrm{g} = 2.25\times10^{-6}\; \mathrm{g}\,
\mathrm{cm}^{-3}$~\citep{2004PhRvL..93b1103K}.

\begin{figure}[!h]
\includegraphics[width=1.0\textwidth]{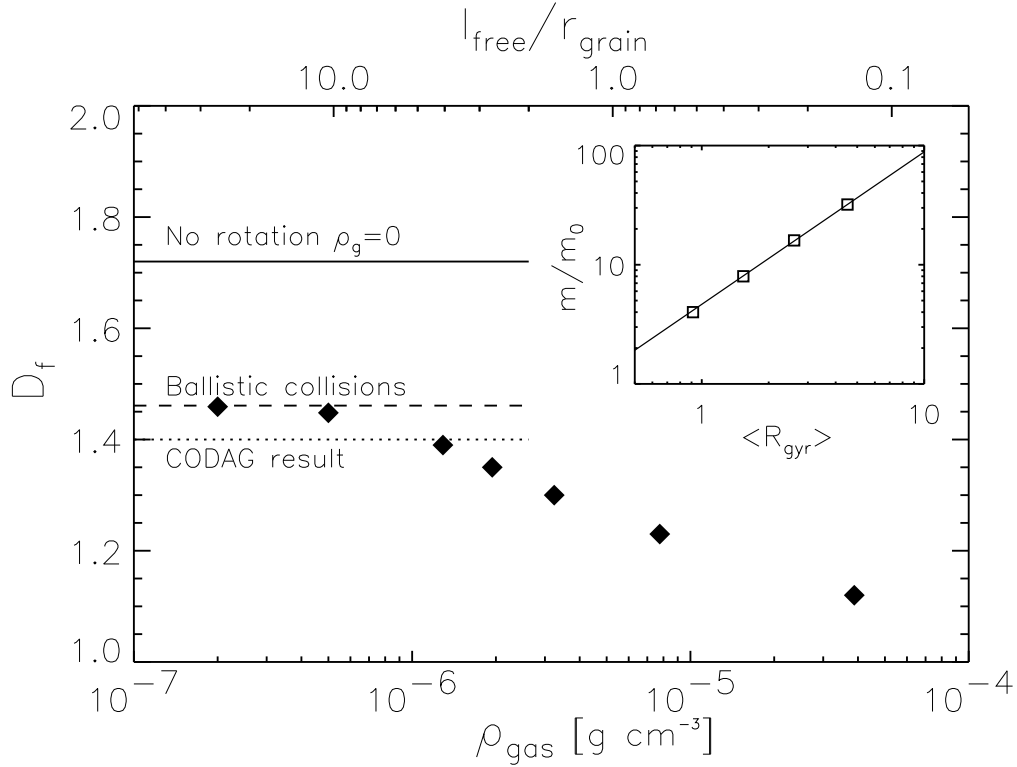}
\caption{ 
Fractal dimension as a function of gas density.  The upper
  x-axis shows the mean free path of a monomer in units of the monomer
  radius.  The horizontal lines show the CODAG measurements
  ($D_\mathrm{f}=1.4$), the limiting result for $\lim_{\rho\to0}
  D_\mathrm{f}=1.46$, and the result for non--rotating aggregates
  $D_\mathrm{f}=1.72$.  The inset plot shows an example of fractal
  dimension determination by slope fitting in an $r_\mathrm{g}$
  versus $m$ plot, at a density of 
  $\rho_\mathrm{g}=3.24\times10^{-6}$.}
\label{df-ro}
\end{figure}

We also calculated the fractal dimension $D_\mathrm{f}$ by fitting a
power law function to the plot of mean aggregate mass versus mean
radius of gyration $r_\mathrm{g}$.  The average was taken over the
entire distribution of particles created for the given mass.  The
fractal dimension as a function of the gas density is shown in
Fig.\ref{df-ro}. For large densities, $D_\mathrm{f}$ appears to follow
a power law with index $\alpha = -0.062$.  Thus for a gas density
$\rho_\mathrm{g}=3.88 \times10^{-5}$ the fractal dimension reaches the
value 1.11.  At low gas densities, the fractal dimension
asymptotically approaches a value of 1.46, slightly larger than
$D_\mathrm{f}=1.41$ as observed in the CODAG experiment. Indeed,
it turns out that the CODAG experiment is operating close to, but not
safely within the ballistic limit.  A gas density of
$2.25\times10^{-6}$\,g\,cm$^{-3}$ is in fact entirely consistent with
$D_\mathrm{f}=1.4$.  At this density, the mean free path of a particle
is only about 1.5 grain diameters.  The experiments therefore are not
in the ballistic limit, but start to feel the influence of
random walk during the collision.  Further increasing the gas
density should strongly enhance this effect, and fractal dimensions
very close to unity should show up when $\rho_\mathrm{g}$ exceeds a
few times $10^{-5} \,\, \mathrm{g\,cm^{-3}}$.

\section{Discussion}

\subsection{The ballistic regime}
In the ballistic limit, rotation increases the collisional
cross-section by producing more opportunities for a contact while two
aggregates are passing each other. In the non--rotation case only
specific 'lucky' parameters (appropriate orientation and impact
parameter) lead to an elongation, while rotation allows a much larger
set of possible collision parameters to be effective.  This effect
leads to more elongated structures, because it
produces significantly more opportunities for collisions with
larger impact parameters.

\subsection{The non-ballistic regime}
In the non--ballistic regime, the random walk executed by the collision
partners during a collision effectively reduces the mean (or
effective) relative translational velocity.  The typical distance
traveled from the original position in a random walk scales with the
square root of the number of steps $N$, we can define an effective
translational speed by
\begin{equation}
<v>=\frac{\sqrt{N}l_\mathrm{b}}{N\tau_\mathrm{f}} \quad .
\label{v_mean_rw}
\end{equation}
If we take $\sqrt{N}l_\mathrm{b}=L$ where $L$ is size of the
aggregate, then the mean effective translational velocity during random
walk is inversely proportional to the gas density $\rho_\mathrm{g}$
(see eq.~\ref{v_mean_rw2},~\ref{tau_f}).
\begin{equation}
<v> = \frac{l_\mathrm{b}^2}{L\tau_\mathrm{f}} \quad .
\label{v_mean_rw2}
\end{equation}
As the gas density increases, the effective velocity decreases and
indeed leads to very long and slow collisions between aggregates.
During that time, the aggregates also execute a random walk in
rotation and in this way expose different parts toward each other. In
the limit of large gas densities, this leads to contact being made
always at maximum distance, and consequently to fully linear
structures. 

It is important to remember that we are only computing collisions of
equal size aggregates, i.e. pure cluster-cluster aggregation (CCA).
This simplification will tend to exaggerate the elongation at a given
size.  If monomers and small aggregates contribute significantly to
the growth, slightly more compact aggregates are produced. In fact, if
a size distribution of impactors is involved in the growth of a
target, the growth physics is an intermediate case between pure CCA
and particle-cluster aggregation (PCA).  As an indication, we can
compare the fractal dimension of 1.7 we found for non--rotating
aggregates in pure CCA with the fractal dimension of $\sim 1.8$
found by \citet{1999Icar..141..388K} for a realistic size distribution.

\subsection{Influence on aggregation timescale}

Elongated aggregates, influenced by rotation and aggregation in
non--ballistic regime, expose a larger surface to the gas and are
easier targets for possible collisions with other dust grains.  Just
like for the particle mass, one can introduce a fractal dimension for
the \emph{cross-section} $D_\mathrm{\sigma}$ and write the relation 
between aggregate shape and cross-section as
\begin{equation}
\sigma \propto r_\mathrm{g}^{D_\mathrm{\sigma}}.
\label{Dsigma}
\end{equation}
Thus for limiting case of compact agglomerates $D_\sigma=2$ while 
for linear grains it is 1.  To see how the crosssection changes 
with aggregate size, one can combine both fractal dimensions 
$D_\mathrm{\sigma}$ and $D_\mathrm{f}$:
\begin{equation}
\sigma \propto (m/m_0)^{\frac{D_\mathrm{\sigma}}{D_\mathrm{f}}}.
\label{SigmaM}
\end{equation}
The cross-section of aggregates is related to aggregation timescale by
\begin{equation}
t \propto \frac{1}{\sigma n v}
\label{agg-tcs}
\end{equation}
where $t$ is aggregation time, $n$ is number density of dust grains 
and $v$ is velocity. We calculated cross-sections of aggregates with 
different fractal dimensions in order to investigate the timescale of 
aggregation. 
\begin{figure}[!h]
\includegraphics[width=1.0\textwidth]{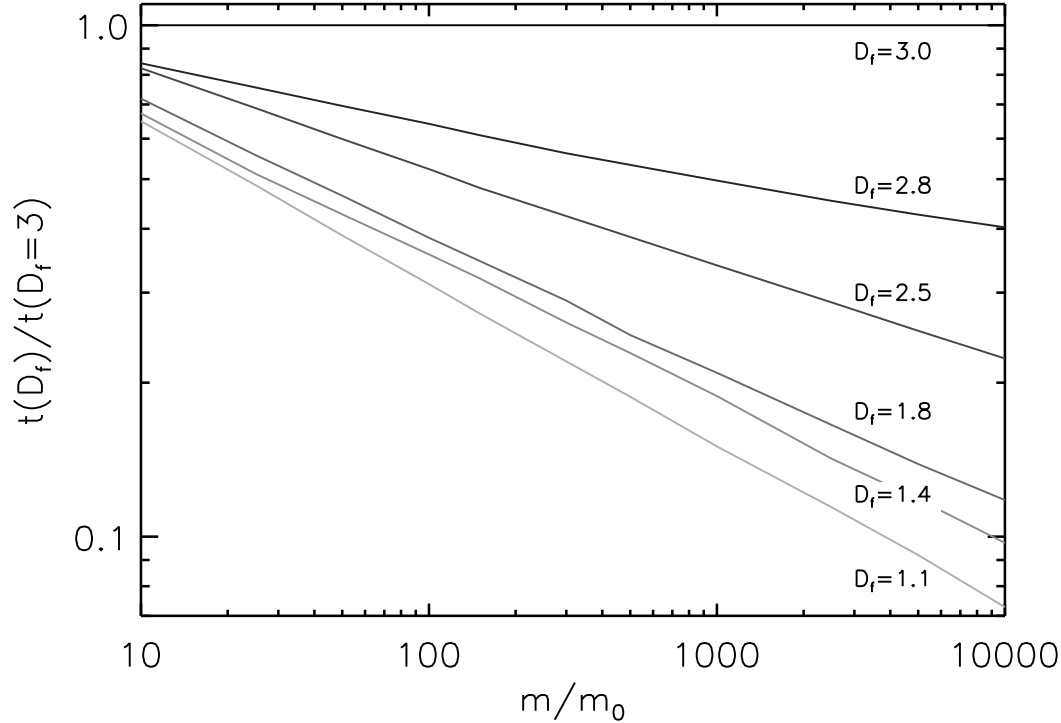}
\caption{ 
  Aggregation timescale $t \propto \frac{1}{\sigma n v}$ divided by
  aggregation timescale for compact agglomerates as a function of
  mass. Different lines correspond to different fractal dimensions.
  From top to bottom $D_\mathrm{f}=3.0,\, 2.8,\, 2.5,\, 1.8,\, 1.4,\,
  1.1.$ }
\label{t-cross}
\end{figure}
Fig.~\ref{t-cross} shows the timescale divided by the timescale for
the most compact aggregates. Each line corresponds to different
fractal dimension $D_\mathrm{f}$. The most interesting are
$D_\mathrm{f}=1.8$ which corresponds to the aggregates formed in the
numerical calculations by \citet{1999Icar..141..388K} and fractal
dimension $D_\mathrm{f}=1.4$ which was obtained in the CODAG
experiment \citep{2004PhRvL..93b1103K}.  The timescale is shorter for
elongated and open aggregates because of the larger cross-sections.
The difference increases with increasing size of aggregates.  

\subsection{Application to protoplanetary disks}

To assess the relevance of the effects discussed in the current paper
in a protoplanetary disk, we need to compute the stopping length of
dust grains as a function of aggregate size and location in the disk.
The boundary condition for ballistic and non--ballistic collisions is
reached when the mean free path equals the physical size of the dust
aggregate approximately given by the radius of gyration:
$l_\mathrm{b}(\rho_\mathrm{g},T)=r_\mathrm{g}(m)$.  This condition
leads to a relation between cluster size and gas density, indicating
how large the cluster should be at a given density in order to start
feeling the effect of elongation enhancement by non--ballistic
collisions.

We will make the simplifying assumption that the stopping time of an
aggregates is (at a given density) independent of size.  This
assumption is valid for very small aggregates, and for aggregates with
very open structure, i.e. low fractal dimension.

We use the  definition of the radius of gyration given by 
eq.~\ref{D_f} and transform it into
\begin{equation}
r_g(m) = B m^{\frac{1}{D_\mathrm{f}}}
\label{r-gyr-df}
\end{equation}
where $B=\frac{1}{A^{1/D_\mathrm{f}}}$ and $A$ is a
proportionality coefficient from eq.~\ref{D_f}.  Then the relation
between cluster mass and gas density relation can be derived as
\begin{equation}
m^{\frac{1}{D_\mathrm{f}}-0.5} = \frac{\varepsilon}{B} 
\sqrt{\frac{3 m_\mathrm{m}}{8 \pi}} \frac{1}{n_0 r^2 \rho_\mathrm{g}},
\label{m-rho}
\end{equation}
where $m_\mathrm{m}$ is a mean mass of a gas molecule and $n_0$ is
number of monomers in the aggregate.  In order to find $n_0$ we
substitute the cluster mass by $m=n_0m_0$ where $m_0$ is the mass of a
monomer.
\begin{equation}
n_0 = \bigg(\frac{\varepsilon}{B} 
\sqrt{\frac{3 m_\mathrm{m}}{8 \pi}}\frac{m_0^{0.5-\frac{1}
{D_\mathrm{f}}}}{r^2 \rho_\mathrm{g}}\bigg)^{\frac{2D_\mathrm{f}}
{2+D_\mathrm{f}}}.
\label{n-rho-eq}
\end{equation}

Eq.~\ref{n-rho-eq} shows how the critical size of a cluster is
dependent on the gas density. It also reveals a dependence on 
monomer size.  We applied eq.~(\ref{n-rho-eq}) to the Hayashi 
model of the solar nebula~\citep{1985prpl.conf.1100H}.  
\begin{figure}[!h]
\includegraphics[width=1.0\textwidth]{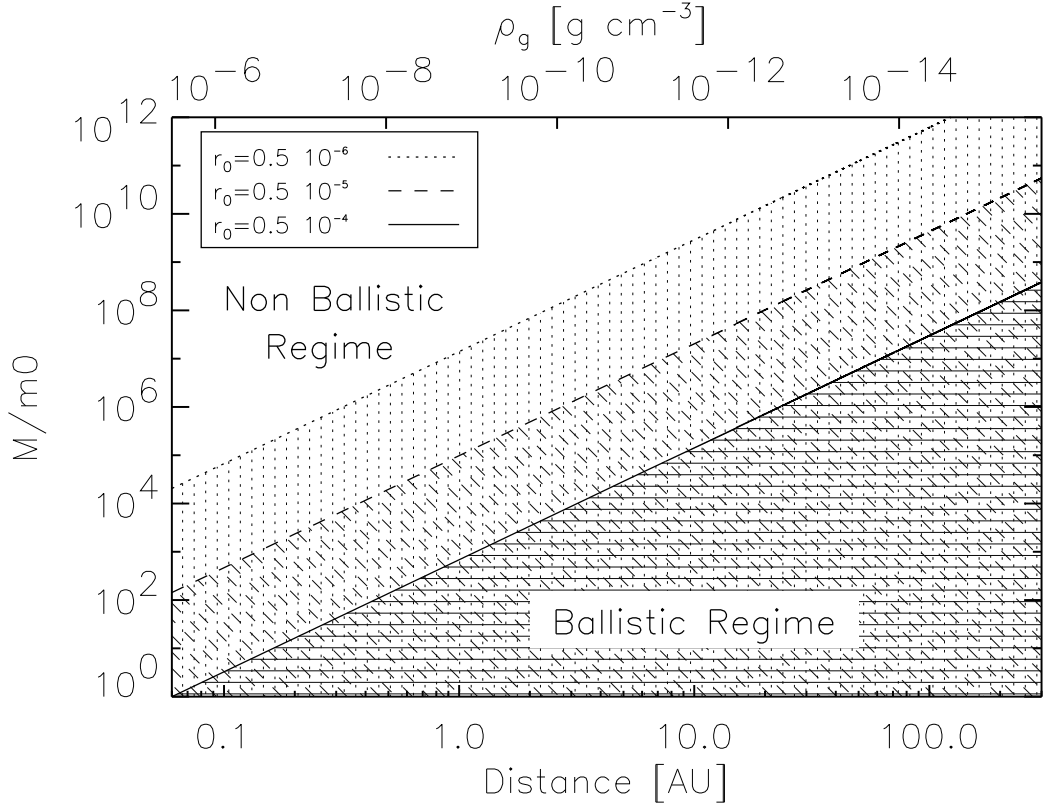}
\caption{ 
  Aggregate mass for which in the midplane of the solar nebula 
  the mean free path $l_\mathrm{free}$ equals the radius of 
  gyration $r_\mathrm{g}$. Different lines correspond to 
  different monomer radii $r_0$.  The dotted line represents a 
  monomer size of $r_0=0.005 \,\mu m$, the dashed line $r_0=0.05 
  \,\mu m$, and the solid line $r_0=0.5 \,\mu m$. The shaded area 
  below each line shows the ballistic regime for corresponding 
  monomer size.  The area above each line represents the 
  non--ballistic regime.  At the top of the diagram, we indicate 
  the midplane density corresponding to the distance from the Sun, 
  according to the Hayashi model~\citep{1985prpl.conf.1100H}.  }
\label{n-rho}
\end{figure}
Fig.~\ref{n-rho} shows, for the midplane density of the model, the
critical size of a cluster as a function of distance from the star.
Each line represents this relation for a different monomer size. 
Thus aggregates with a fractal dimension of $D_\mathrm{f}=1.46$, 
are formed below the line, while above that line more elongated 
grains are produced.  Each line is a border between the ballistic 
and non--ballistic regimes for a given monomer size.  For example, 
at a distance below 1 AU from the star, the gas density strongly 
influences the shape of aggregates formed through Brownian motion, 
if the aggregates consist of more than a few tens of 1$\mu$m grains.
Consequently, the CODAG experiment reproduced conditions present in
the inner part of the protoplanetary disk while its applicability to
the low density regions in the outer disk is limited.  At the CODAG
density of $\rho_\mathrm{g}=2.25 \times
10^{-6}\,\mathrm{g}\,\mathrm{cm}^{-3}$, the critical size is about 1.5
micron sized monomers.

\section{Conclusions}

We have studied the effect of aggregate rotation during collisions.
The results show that rotation must definitely be treated
correctly when modeling growth due to aggregation, or the 
geometrical structure of the resulting aggregates will be incorrect.  

Rotation plays a role because it enhances the probability that two
approaching aggregates make contact early on during the collision, 
between outer constituent grains. For the most simple case of 
ballistic collisions, during which the colliding aggregates move on 
a linear path, ignoring aggregate rotation increases the fractal 
dimension from 1.46 to 1.7 in the limit of pure cluster-cluster 
aggregation.

This effect becomes strongly enhanced if the density of the
surrounding medium becomes so large that the stopping length of an
aggregate becomes shorter than the aggregate size, i.e. in the
non--ballistic limit.  We find that reducing the stopping length to 
the monomer radius results in a fractal dimension of 1.25.  When the
stopping length is reduced to one tenth of the monomer radius, the
fractal dimension drops to 1.1.

For the solar nebula, we find that non--ballistic collisions play a
role in the innermost regions of the disk for even very small
aggregates made of a few $\mu m$-sized grains.  For smaller monomers
and/or further away from the Sun, enhanced elongation can be expected
for larger aggregates, made of a few 100 or 1000 grains. In the outer
disk, all collisions may be considered ballistic.

\section*{Acknowledgments}
This work was supported by the Nederlandse Organisatie voor
Wetenschapelijk Onderzoek, Grant 614.000.309.  We thank J. Blum for 
an insightful referee report, and for providing data obtained in
CODAG microgravity experiment.  We also thank the second (anonymous)
referee for useful comments that lead to a restructuring of the
manuscript.

\bibliographystyle{elsart-harv}
\bibliography{citation}

\begin{thebibliography}{14}
\expandafter\ifx\csname natexlab\endcsname\relax\def\natexlab#1{#1}\fi
\expandafter\ifx\csname url\endcsname\relax
  \def\url#1{\texttt{#1}}\fi
\expandafter\ifx\csname urlprefix\endcsname\relax\def\urlprefix{URL }\fi

\bibitem[{{Blum} and {Wurm}(2000)}]{2000Icar..143..138B}
{Blum}, J., {Wurm}, G., Jan. 2000. {Experiments on Sticking, Restructuring, and
  Fragmentation of Preplanetary Dust Aggregates}. Icarus 143, 138--146.

\bibitem[{{Blum} et~al.(1996){Blum}, {Wurm}, {Kempf}, and
  {Henning}}]{1996Icar..124..441B}
{Blum}, J., {Wurm}, G., {Kempf}, S., {Henning}, T., Dec. 1996. {The Brownian
  Motion of Dust Particles in the Solar Nebula: an Experimental Approach to the
  Problem of Pre-planetary Dust Aggregation}. Icarus 124, 441--451.

\bibitem[{{Blum} et~al.(2000){Blum}, {Wurm}, {Kempf}, {Poppe}, {Klahr},
  {Kozasa}, {Rott}, {Henning}, {Dorschner}, {Schr{\" a}pler}, {Keller},
  {Markiewicz}, {Mann}, {Gustafson}, {Giovane}, {Neuhaus}, {Fechtig}, {Gr{\"
  u}n}, {Feuerbacher}, {Kochan}, {Ratke}, {El Goresy}, {Morfill},
  {Weidenschilling}, {Schwehm}, {Metzler}, and {Ip}}]{2000PhRvL..85.2426B}
{Blum}, J., {Wurm}, G., {Kempf}, S., {Poppe}, T., {Klahr}, H., {Kozasa}, T.,
  {Rott}, M., {Henning}, T., {Dorschner}, J., {Schr{\" a}pler}, R., {Keller},
  H.~U., {Markiewicz}, W.~J., {Mann}, I., {Gustafson}, B.~A., {Giovane}, F.,
  {Neuhaus}, D., {Fechtig}, H., {Gr{\" u}n}, E., {Feuerbacher}, B., {Kochan},
  H., {Ratke}, L., {El Goresy}, A., {Morfill}, G., {Weidenschilling}, S.~J.,
  {Schwehm}, G., {Metzler}, K., {Ip}, W.-H., Sep. 2000. {Growth and Form of
  Planetary Seedlings: Results from a Microgravity Aggregation Experiment}.
  Physical Review Letters 85, 2426--2429.

\bibitem[{{Boss}(1997)}]{1997Sci...276.1836B}
{Boss}, A.~P., 1997. {Giant planet formation by gravitational instability.}
  Science 276, 1836--1839.

\bibitem[{{Cuzzi} et~al.(2001){Cuzzi}, {Hogan}, {Paque}, and
  {Dobrovolskis}}]{2001ApJ...546..496C}
{Cuzzi}, J.~N., {Hogan}, R.~C., {Paque}, J.~M., {Dobrovolskis}, A.~R., Jan.
  2001. {Size-selective Concentration of Chondrules and Other Small Particles
  in Protoplanetary Nebula Turbulence}. \apj 546, 496--508.

\bibitem[{{Dominik} and {Tielens}(1997)}]{1997ApJ...480..647D}
{Dominik}, C., {Tielens}, A.~G.~G.~M., May 1997. {The Physics of Dust
  Coagulation and the Structure of Dust Aggregates in Space}. \apj 480, 647--+.

\bibitem[{{Draine} and {Weingartner}(1996)}]{1996ApJ...470..551D}
{Draine}, B.~T., {Weingartner}, J.~C., Oct. 1996. {Radiative Torques on
  Interstellar Grains. I. Superthermal Spin-up}. \apj 470, 551--+.

\bibitem[{{Hayashi} et~al.(1985){Hayashi}, {Nakazawa}, and
  {Nakagawa}}]{1985prpl.conf.1100H}
{Hayashi}, C., {Nakazawa}, K., {Nakagawa}, Y., 1985. {Formation of the solar
  system}. In: Protostars and Planets II. pp. 1100--1153.

\bibitem[{{Kempf} et~al.(1999){Kempf}, {Pfalzner}, and
  {Henning}}]{1999Icar..141..388K}
{Kempf}, S., {Pfalzner}, S., {Henning}, T.~K., Oct. 1999.
  {N-Particle-Simulations of Dust Growth. I. Growth Driven by Brownian Motion}.
  Icarus 141, 388--398.

\bibitem[{{Krause} and {Blum}(2004)}]{2004PhRvL..93b1103K}
{Krause}, M., {Blum}, J., Jul. 2004. {Growth and Form of Planetary Seedlings:
  Results from a Sounding Rocket Microgravity Aggregation Experiment}. Physical
  Review Letters 93~(2), 021103--+.

\bibitem[{{Ossenkopf}(1993)}]{1993A&A...280..617O}
{Ossenkopf}, V., Dec. 1993. {Dust coagulation in dense molecular clouds: The
  formation of fluffy aggregates}. \aap 280, 617--646.

\bibitem[{{Purcell}(1979)}]{1979ApJ...231..404P}
{Purcell}, E.~M., Jul. 1979. {Suprathermal rotation of interstellar grains}.
  \apj 231, 404--416.

\bibitem[{{Weidenschilling}(1980)}]{1980Icar...44..172W}
{Weidenschilling}, S.~J., Oct. 1980. {Dust to planetesimals - Settling and
  coagulation in the solar nebula}. Icarus 44, 172--189.

\bibitem[{{Youdin} and {Shu}(2002)}]{2002ApJ...580..494Y}
{Youdin}, A.~N., {Shu}, F.~H., Nov. 2002. {Planetesimal Formation by
  Gravitational Instability}. \apj 580, 494--505.

\end{thebibliography}
\end{document}